\documentclass[prd,eqsecnum,tightenlines,11pt]{revtex4}

\usepackage{hyperref}
\usepackage{graphicx}
\usepackage{amsmath,amssymb,amsfonts,amsthm,latexsym,stmaryrd}
\usepackage{marginnote}
\usepackage{color}
\usepackage{soul}

\begin{document}

\title{Scalar quasinormal modes of gravastars}
\author{Hamza Boumaza}
\affiliation{\small Laboratoire de Physique des Particules et Physique Statistique (LPPPS),\\
 Ecole Normale Supérieure-Kouba, B.P. 92, Vieux Kouba, 16050 Algiers, Algeria}

\begin{abstract}
{\hskip 2em}In this work, we present a detailed investigation of gravastars within the framework of scalar-tensor theories, emphasizing both the background and perturbed levels for trivial and non trivial scalar field.  We derive and analyze the background equations governing the equilibrium configurations of gravastars by considering the interaction between matter and scalar field through a conformal transformation.  By selecting two forms of the coupling function, we identify distinct physical characteristic that leads to deviations from predictions made by general relativity. We show that  that the scalar field  has a considerable influence on the compactness parameter. By extending our study to the perturbed level, we show also that quasinormal modes are considerably impacted by the presence of scalar field. New and interesting solutions are found, which are  absent in general relativity.
 \end{abstract}

\maketitle

\section{Introduction:}
Although the  black holes (BHs), which are considered as a solution to Einstein's equations in the vacuum, are perfectly compatible with gravitational wave \cite{abbott2016virgo} and  are accepted by astrophysical researcher, their nature and their physical existence is  one of the most question in theoretical astrophysics. The scepticism regarding their physical reality   arises from the exotic solution of Einstein's equations. In other words, the presence of a central singularity and an event horizon at the Schwarzschild radius makes the physical proprieties of classical black hole unknown.  The  recent observation by the Event Horizon Telescope (ETH) of the supermassive black hole (M87) \cite{akiyama2019first} confirms the existence of such object and open new window to probe the interior of BHs. It is  plausible to expect that there could be horizonless compact objects as alternatives to BHs which have been extensively explored in the literature, for example the Proca star \cite{brito2016proca}, Boson stars \cite{schunck2003general}, and gravastars (gravitational vacuum stars) \cite{mazur2004gravitational,mazur2023gravitational,carter2005stable}.\\

 In General Relativity (GR) framework, gravastars  was proposed as a solution to the paradoxes associated with event horizons where  the interior solution of a black hole is supposed to be de Sitter space. The simplest theoretical construction of gravastars is: de Sitter  spacetime core, with an equation of state $p=-\rho$,  connected to a Schwarzschild spacetime exterior through a thin shell of perfect-fluid matter. Many papers have contributed to understand gravastars within the broader context of exotic compact objects since it was introduced by Mazur and Mottola  \cite{mazur2004gravitational} for the first time. Instability against radial perturbation was investigated in Ref.\cite{carter2005stable}, where they found a large parameters that gives stable gravastars. In Ref.\cite{chirenti2007tell}, the authors  computed the axial oscillation modes  of gravastars and they did not only find any instabilities but they also concluded that one can make a difference between a gravastar and a black hole through the Quasi-Normal Modes (QNMs). Pani et al. \cite{pani2009gravitational,pani2010gravitational} reached similar conclusions by analyzing the polar and axial perturbation equations, taking into account the junction conditions that connect the internal and external regions of gravastars. Their works highlighted the important role of compactness and surface properties, which have significant astrophysical implications on the behavior of these exotic objects.\\
 
 In this paper, we aim to investigate  gravastars in the framework of tensor-scalar field theory due to their potential to prob modified gravitational effects near the surface. These theories are extension to general relativity where a scalar field is introduced  as an extra degree of freedom, which allow us explore gravitational field in strong regimes and provide  behaviors different from GR \cite{fujii2003scalar,Berti:2015itd,doneva2020stability}.  The scalar field seems to have an important astronomical implications in the  acceleration of the late universe \cite{Boumaza:2020klg,Langlois:2018dxi,Langlois:2017dyl} as well as  modifying the features of compact object such as: black holes \cite{BenAchour:2020wiw,Minamitsuji:2019shy,Motohashi:2019sen} and neutron stars \cite{Boumaza:2021fns,Ogawa:2019gjc,Boumaza:2022abj,Babichev:2016jom,Boumaza:2021lnp,cisterna2015neutron,cisterna2016slowly}. Moreover, in some subclass of these theories, an interesting  phenomenon called \textit{spontaneous scalarization} \cite{Damour:1996ke,silva2018spontaneous} occurs in which the scalar field   arises from a linear tachyonic instability around  compact objects \cite{doneva2024spontaneous,mendes2016highly}. The scalarization  phenomenon  with different models  has attract a lot of attention lately\cite{minamitsuji2021spontaneous,boumaza2023axial,Pani:2014jra,kuan2021dynamical,yazadjiev2016slowly,doneva2018neutron,andreou2019spontaneous} for two features. The first is the GR solution that we observe for a trivial scalar field and for weak gravitational regime. The second  is the phase transition from a trivial scalar field to a nontrivial onewhen the mass (for BHs)  or compactness (for neutron stars) exceeds a certain interval. Consequently, depending on the coupling function of the scalar field with the metric, a deviation of the  solutions in scalar tensor theories  from GR will be noticed during the transition.\\

  Motivated by the scalarezation phenomenon, we restrict our  intention to the physical characteristic of gravstars with a constant energy density coupled to the matter field through a conformal transformations \cite{Bekenstein:1992pj}. Quasinormal modes, which  are intrinsic oscillations  characterized by complex frequencies that encode both oscillation and damping,  will play a pivotal role in gravitational wave spectroscopy allowing us to extract the gravastars properties from observed waveforms. Studying QNMs is crucial for understanding the stability and resonant behavior of the scalarezation phenomenon as well as the exotic compact objects \cite{benhar1999imprint,kokkotas1999quasi,kokkotas1992w,motahar2019axial}. Furthermore, these modes can be a useful to examine the interaction between the scalar field and the exotic matter and to assess the feasibility of such objects in the universe dominated by dark energy.\\
  
 The structure of this paper is as follows: In Section \ref{sec1}, we provide a brief review of the matter distribution of gravastars and the underlying structure of spacetime. Section \ref{sec2} introduces the background and perturbation equations that profile our exotic object within the framework of scalar-tensor theory. In Section \ref{sec3}, we derive the analytic solution for the perturbed scalar field equation in the interior of the star, assuming a trivial scalar field. Additionally, we compute the QNMs using the continued fraction method. Section \ref{sec4} presents a detailed numerical analysis of our findings. Finally, the paper concludes with a summary and discussion in the concluding section.

\section{Brief Review of gravastars in general relativity:}\label{sec1}
Let us start by presenting the main layers that define a gravastars. In Ref.\cite{mazur2004gravitational}, the authors divide the star into three regions where the matter is distributed in the spacetime with three different equation of states:
\begin{itemize}
 \item[1.]Interior, $r<a$, with $p=-\rho$.
 \item[2.]Thin shell, with a surface tension $\Theta$ and a vanishing surface density.
 \item[3.]Exterior, $r>a$, with $\rho=p=0$.
\end{itemize}

where $p$ and $\rho$ are the pressure and the energy density of the fluid and $a$ is the radius of the star. It is obvious  that the interior energy density can be written as $\rho=3M/(4\pi a^3)$, where $M$ is the total mass. In a static spherically symmetric metric,
\begin{eqnarray}\label{ds}
ds^2=-f(r)dt^2+h(r)dr^2+r^2(d\theta^2+\sin\theta^2 d\phi^2),
\end{eqnarray}
the equation of gravitational field for a gravastar are given by
\begin{eqnarray}
\frac{f'(r)}{f(r)}&=&-8 \pi  r h(r) \rho +\frac{h(r)-1}{r},\\
\frac{h'(r)}{h(r)}&=& 8 \pi  r h(r) \rho +\frac{1-h(r)}{r}.
\end{eqnarray}
Integrating these equations and imposing the regularity at the center, the solutions reads
\begin{eqnarray}\label{grbacground}
f(r)=1/h(r)=
\left\{
        \begin{array}{ll}
            1-\frac{8\pi\rho a}{3}r^2,&\qquad r<a\\
            &\\
            1-\frac{r_s}{r},&\qquad r>a
        \end{array}
    \right.
\end{eqnarray}
where the schwarzschild radius, $r_s$, is less than the radius $a$ to avoid the singularity. By replacing the EoS of the interior fluid in the matter continuity equation which reads
\begin{eqnarray}
p'=-\frac{f'(r)}{2f(r)}(p+\rho),
\end{eqnarray} 
give us  a constant pressure and thus a constant energy density. By imposing the continuity at the surface of the object, we find
\begin{eqnarray}\label{rs=M}
r_s = 2 M.
\end{eqnarray}
Thus, as long as we have $r_s<a$, the singularity is avoided and  when we take the limit $r_s\rightarrow a$ we get the model discussed in Ref.\cite{mazur2004gravitational}. However, the first derivative is discontinuous at the surface which is due to the junction conditions \cite{Israel:1966rt}
\begin{eqnarray}\label{junctionconditions}
[[\nabla_{i}n_{j}]]=8\pi [[\Theta u_{i}u_{j}]],\qquad with\quad \{i,j\}=\{t,\theta,\phi\},
\end{eqnarray}
where $n_{\alpha}=(0,1/\sqrt{\tilde{h}(r)},0,0)$ and $u_{\alpha} =(1/\sqrt{\tilde{f}(r)},0,0,0)$ are the unite normal vector and the four-velocity vector.The notation [[A]] is the discontinuity of "A" across the thin shell. Inserting eq.(\ref{ds}) into eq.(\ref{junctionconditions}), we get
\begin{eqnarray}
8\pi\Theta a= -\frac{3r_s/a}{2\sqrt{1-r_s/a}}.
\end{eqnarray}
As we can see a gravastar can be characterized only by its compactness $C =r_s/a$. Using the transformation
\begin{eqnarray}\label{coordinatchangment}
r\, \rightarrow a \,\hat{r},
\end{eqnarray}
the background solution becomes
\begin{eqnarray}\label{grbacgroundre}
f(\hat{r})=1/h(\hat{r})=
\left\{
        \begin{array}{ll}
            1-C\,\hat{r}^2,&\qquad \hat{r}<1,\\
            &\\
            1-\frac{C}{\hat{r}},&\qquad \hat{r}>1.
        \end{array}
    \right.
\end{eqnarray}
Therefore, the radius of the star is not important in our analysis since the equations and their solutions are independents of $a$. As we will see in the next section, this redefinition is going to simplify our work even when extend our model by in introducing an extra degree of freedom scalar field.

\section{Equations of motions of scalarized gravastar:}\label{sec2}
In this section, we present the theoretical framework underpinning this study.  We will introduces the main principles and assumptions that define the model, detailing the specific variables, parameters, and interactions. Following this, we elaborates on the mathematical formulation of the model, presenting the key equations governing its dynamics and behavior. Starting with the background equations that describe the fundamental, unperturbed state, we then introduce the perturbed equations, which account for deviations from the background state, allowing us to explore small fluctuations and their impact on the gravastar's behavior.

\subsection{The model}
Now, we propose to study the gravastar presented in the last section in the presence of a scalar field $\varphi$. We will consider the model  
\begin{eqnarray}\label{action}
S=\int d^4x \sqrt{-g} \left(\frac{1}{16\pi}R-\partial_\mu\varphi\partial^\mu\varphi\right)+\int d^4x \sqrt{-\tilde{g}} \tilde{p}.
\end{eqnarray}
where $R$ is the Ricci scalar and  the tilde denotation is used to express any quantity in Jordane frame. The first integral of the action, which describes the spacetime, is written in Einstein frame while the second one  the metric is in Jordane frame. The relation between the two frames is expressed through the conformal transformation
\begin{eqnarray}
\tilde{g}_{\mu\nu}=\text{A}(\varphi)g_{\mu\nu},
\end{eqnarray}
where $\text{A}(\varphi)$ is an arbitrary function of $\varphi$. Indeed, the matter interacts with the scalar field via this function, which is expected to alter the metric's behavior and influence the star's characteristics. Without loss of generality, by considering that the value of the scalar field at the present time is zero, one can  make a Taylor development of $\alpha(\varphi)=d\log(\text{A}(\varphi))/d\varphi$ around $0$ as 
\begin{eqnarray}
\alpha(\varphi)=\alpha_0 + \beta_0\, \varphi^2+ O[\varphi^4],
\end{eqnarray}
where
\begin{eqnarray}
\beta(\varphi)=\frac{d\alpha(\varphi)}{d\varphi}.
\end{eqnarray}
In order to satisfy the constraint made by solar-system experiment $\alpha_0<10^{-3}$ \cite{Bertotti:2003rm}, we set $\alpha_0=0$ \cite{Damour:1996ke,Damour:2009vw,damour1993nonperturbative}. The constant $\beta_0$ is crucial to have spontaneous scalarization \cite{Mendes:2016fby}, where if it is sufficiently negative, we expect an evolution of scalar field which will change the metric solution of GR to an other solution with non-trivial scalar field. In fact, a neutron star becomes unstable when we increase the compactness and it undergoes a phase transition to a scalarized configuration \cite{berti2015testing,Mendes:2016fby}. An other characteristic, we should impose is the reflection symmetry $\varphi\rightarrow - \varphi$ to our model, thus it is convenient to choose the following functions
\begin{eqnarray}\label{models}
\text{M}_1:& \text{A}(\varphi)&=e^{\beta\,\varphi^2},\\
\text{M}_2:& \text{A}(\varphi)&=(\cosh(\sqrt{6}\beta\,\varphi^2))^{\frac{1}{3\beta}}.
\end{eqnarray}
Thus, we determine that $\beta=\beta_0$ and $\alpha=\alpha_0$ in those models. Both models are identical under first- and second-order perturbations, but they differ at higher orders.  The first model is the most commonly used in the literature because it is the simplest function that gives the spontaneous  scalarization effect \cite{Doneva:2022ewd,Berti:2015itd}. The model $\text{M}_2$ is a good approximation  of the action (\ref{action})  equivalent to the action with a massless  scalar field $\Phi$ nonminimally coupled to gravity
\begin{eqnarray}\label{action2}
S=\int d^4x \sqrt{-g} \left(\frac{1}{16\pi}(1-8\pi\xi\Phi^2)\tilde{R}-\partial_\mu\Phi\partial^\mu\Phi\right)+\int d^4x \sqrt{-\tilde{g}} \tilde{p}.
\end{eqnarray}
where  $\xi$ is a real constant called the conformal coupling parameter and 
\begin{eqnarray}
\frac{d\varphi}{d\Phi}&=&4\pi \frac{\sqrt{1-8\pi\xi(1-6\xi)\Phi^2}}{1-8\pi\xi\Phi^2},\\
\text{A}(\varphi)&=&\frac{1}{\sqrt{1-8\pi\xi\Phi^2}}.
\end{eqnarray}
At small regime $\varphi\rightarrow 0$ and $\Phi\rightarrow 0$, one can identify that $\xi=\beta/2$ by integrating the above equation. Note that the theory described by (\ref{action2}) provides an exponential  growth of the scalar field in the presence of gravitational fields produced by relativistic matter through a quantum mechanism \cite{Landulfo:2014wra}.

\subsection{The equations of motions}
To derive the background field equations, we insert the metric (\ref{grbacground})  and the scalar field $\varphi\equiv \varphi (r)$ in (\ref{action}), and then we vary the equations with respect to $f$, $h$ and $\varphi$. Doing so, we get
\begin{eqnarray}
\frac{h'(r)}{h(r)}&=&\frac{1-h(r)}{r}+8 \pi  r  \varphi '(r)^2+8 \pi  r h(r)  \text{A}(\varphi (r))^2 \rho, \label{eh} \\
\frac{f'(r)}{f(r)}&=&\frac{h(r)-1}{r}+8 \pi  r  \varphi '(r)^2-8 \pi  r h(r)  \text{A}(\varphi (r))^2\rho, \label{ef} \\
\varphi''(r)&=& \varphi '(r) \left(\frac{h'(r)}{2 h(r)}-\frac{f'(r)}{2 f(r)}-\frac{2}{r}\right)+h(r)  \text{A}(\varphi (r)) \text{A}_{\varphi}(\varphi (r))\rho.\label{ephi}
\end{eqnarray}
where the prime denote the derivative with respect to the coordinate $r$ and the subscribe $\varphi$ the derivative with respect to $\varphi$. By examining the above equations, we can see that the metric (\ref{grbacground}) is obtained for a vanishing scalar field, is secured by the requirement
\begin{eqnarray}\label{condition}
\text{A}_{\varphi}(0)=0.
\end{eqnarray}
The coupling functions of the models $\text{M}_1$ and $\text{M}_2$ satisfy this condition, which means that GR gravastar is a  solution within scalar tensor theories gravity. It is difficult to get the general analytic expressions of the metric and scalar field, but one can integrate the equations numerically and thus obtain the characteristics of the star.\\

Now, we investigate the perturbed equations of the first order by considering the perturbations $g_{\mu\nu}=g_{\mu\nu}^{(0)}+\delta g_{\mu\nu}$ and $\varphi=\varphi(r)+\delta\varphi(t,r)$, where $g_{\mu\nu}^{(0)}$ is the background metric and
\begin{eqnarray}
\delta g_{\mu\nu}=\
\left(
\begin{array}{cccc}
 f(r) H_0^{lm} & \sqrt{f(r)} \sqrt{h(r)} H_1^{lm} & 0 & 0 \\
 \sqrt{f(r)} \sqrt{h(r)} H_1^{lm}& h(r) H_2^{lm} & r \sqrt{h}H_5^{lm} \partial_\theta Y_{lm}& r\sin\theta\sqrt{h}H_5^{lm} \partial_\phi Y_{lm} \\
 0  &r \sqrt{h}H_5^{lm} \partial_\theta Y_{lm}  & 0 & 0 \\
0  & r\sin\theta\sqrt{h}H_5^{lm} \partial_\phi Y_{lm}  & 0 & 0 \\
\end{array}
\right).
\end{eqnarray} 

Here we chose the uniform curvature gauge. $H_0^{lm}$, $H_1^{lm}$, $H_2^{lm}$ and $H_5^{lm}$ depend on the coordinates $t$ and $r$, and $Y_l^m(\theta,\phi)$ are the spherical harmonic functions. For simplicity, we are going  to  remove the (sub/super)script "$lm$" from the perturbed quantities, since they  will  not contribute to the perturbed equations. By considering the second order expansion of the total action (\ref{action}) and after integrating by parts, it follows that 
\begin{eqnarray}\label{S2polar}
\delta^{2}S^{polar} &=&\int drdt\left(H_0\left(a_1 \delta\varphi' +L a_2 H_5' +L a_3 H_2'+a_4 \delta\varphi +a_5  H_5 +a_6 H_2\right)+L a_7 H_1^2\right.\nonumber\\
 &&\left. +H_1(a_8\delta\dot{\varphi} +L a_{9} \dot{H}_5 + a_{10} \dot{H}_2)+ a_{11} H_2^2+H_2(a_{12}\delta\varphi' +a_{13}\delta\varphi +L a_{14}H_5 )  \right.\nonumber\\ 
 &&\left.+ L a_{15}H_5^2 + L a_{16}\dot{H}_5^2 + L a_{17}H_5 \delta\varphi + e_1 \delta\varphi^2 + e_2 \delta\varphi'^2 + e_3 \delta\dot{\varphi}^2\right).
\end{eqnarray}
where $L=l(l+1)$ and the coefficients $a_i$ and $e_i$ are given in  appendix  \ref{app.A}. To eliminate the non-dynamical variables, we must distinct three cases depending on the integer $l$.
\subsubsection{The case $l \geq 2$:}\label{lsup2}
We vary (\ref{S2polar}) with respect to $H_0$ and $H_1$, respectively, to obtain
\begin{eqnarray}
&& a_1 \delta\varphi' +L a_2 H_5' + a_3 H_2'+a_4 \delta\varphi +a_5  H_5 +a_6 H_2=0,\\
&& 2 L a_7 H_1+a_8\delta\dot{\varphi} +L a_{9} \dot{H}_5 + a_{10} \dot{H}_2=0,
\end{eqnarray}
and we also  define  the combination \cite{DeFelice:2011ka,Kobayashi:2014wsa,kase2020stability,minamitsuji2016relativistic}
\begin{eqnarray}
&& \psi = L a_2 H_5 + a_3 H_2.
\end{eqnarray}
The last three  equations are solved with respect to $H_1$, $H_2$ and $H_5$. Then, substituting the obtained solutions  in (\ref{S2polar}), the time derivative of $H_5$ and $H_2$ are eliminated from the action. Therefore, the only functions left in (\ref{S2polar}) are  $\psi$ and $\delta \varphi$. After long calculations, we arrive to
\begin{eqnarray}\label{deltaS2}
\delta^{2}S &=& \int drdt\;\left( \dot{\chi}^t\textbf{K}\dot{\chi}+{\chi^{t}}' \textbf{G}\chi' +\chi^t \textbf{L}\chi' +\chi^t \textbf{M}\chi\right),
\end{eqnarray}
with
\begin{eqnarray}
{\chi}^t=\{\psi,\delta \varphi\},
\end{eqnarray}
and $\textbf{K}$, $\textbf{G}$ and $\textbf{M}$ are $2\times 2$ matrices, where their expression are complicated, but they are reduced considerably in the case $\varphi=0$ and under the condition $A_\varphi(0)=0$.

\subsubsection{The case $l=0$:}

 If we impose  $l=0$ in the action (\ref{S2polar}), we must reduce the degrees of freedom by choosing the gauge $H_0=0$ (or $H_1=0$). Therefore, the action (\ref{S2polar}) is reduced to
\begin{eqnarray}
\delta^{2}S &=& \int drdt \left(H_1(a_8\delta\dot{\varphi} + a_{10} \dot{H}_2)+ a_{11} H_2^2+H_2(a_{12}\delta\varphi' +a_{13}\delta\varphi ) +  e_1 \delta\varphi^2 + e_2 \delta\varphi'^2 + e_3 \delta\dot{\varphi}^2 \right).\nonumber\\ 
\end{eqnarray}
The equation, which is derived from the variation of this action with regard to $H_1$, can be solved for $\dot{H}_2$. By substituting the result  back into the action, we reformulate it as 
\begin{eqnarray}
\delta^{2}S&=& \int drdt\;\left( e_3 \delta\varphi^2 + e_2 \delta\varphi'^2 + e_1 \delta\dot{\varphi}^2 \right),
\end{eqnarray} 
with
\begin{eqnarray}
 \tilde{e}_3&=&  \tilde{e}_3 +\frac{a_{11}a_{8}^2}{a_{10}^2}-\frac{a_{11}a_{8}}{a_{10}}+\frac{d}{dr}\left[\frac{a_{12}a_{8}}{a_{10}}\right].
\end{eqnarray}
When $\varphi=0$ the equation of the perturbed scalar field is a particular case of the equation derived from (\ref{deltaS2}), by taking $l=0$.
\subsubsection{The case $l=1$:}
For $l=1 $, it is necessary to constrain the degree of freedom by establishing the gauge condition$ \delta\varphi=0 $.  Following the same steps for the case $l\geq 2$, we obtain
\begin{eqnarray}
\delta^{2}S&=& \int drdt\;\left( \tilde{a}_3 \delta\psi^2 + \tilde{a}_2 \delta\psi'^2 + \tilde{a}_1 \delta\dot{\psi}^2 \right).
\end{eqnarray} 
We calculate that the total action vanish when $\varphi$ is constant. Therefore, in this paper, we will not take into consideration the case $l=1$.

\section{Exact solutions  for vanishing scalar field:}\label{sec3}
In this section, we suppose that $\varphi =0 $ which reduce the gravastars of GR case for the condition (\ref{condition}) and thus one can study the the first order perturbed scalar field independently from the metric. Since, in our work, we are interested by the scalar field propagation, we will  only study scalar field's equations. Varying the action (\ref{deltaS2}) w.r.t. $\delta\varphi$ and using the Fourier transform $\delta\varphi\rightarrow\int dt\, e^{I\omega t } \delta\varphi/r$, we obtain
\begin{eqnarray}\label{equper}
\frac{\partial^2 \delta\varphi}{\partial r^{*2}}+(\omega^2 -V(r) )\delta\varphi=0,
\end{eqnarray}
where $r^*$ is the tortoise coordinate and $V(r)$ has the  form
\begin{eqnarray}
V(r)= \frac{f \rho  \left(\kappa  A_{\varphi\varphi}(0 )-1\right)}{\kappa }+\frac{f (h (l(l+1)+1)-1)}{h r^2},
\end{eqnarray}
where we  chose that the  conformal function  verifies $A(0)=1$. For the both models $\text{M}_1$ and $\text{M}_2$, we remarque that  $2\beta =A_{\varphi\varphi}(0 )$ which means  they are indistinguishable for the vanishing scalar field case. However, the both model differ from the GR case where the constant $\beta$ will play an important role to have a deviation from GR. The explicit form of the potential $V(r)$ is given by
\begin{eqnarray}
V(\hat{r})=\left\{
        \begin{array}{ll}
            \frac{l(l+1)}{\hat{r}^2}-2 C^2 \hat{r}^2 (3 \beta  \kappa -1)+C (l(l+1)+2-6 \beta  \kappa ),&\qquad \hat{r}<1,\\
            &\\
            -\frac{C^2}{\hat{r}^4}+\frac{C(1- l (l+1))}{\hat{r}^3}+\frac{l(l+1)}{\hat{r}^2},&\qquad \hat{r}>1.
        \end{array}
    \right.
\end{eqnarray}
The standard method for deriving the interior solution for perturbed stars is the direct integration of the  ordinary differential equation (\ref{equper}) from the center of the star to its radius. In the de Sitter interior ($r < 1$), numerical integrations are unnecessary, as a regular solution to the perturbation equations may be expressed using hypergeometric functions. To have a physical  solution, we require regularity everywhere inside the star. At the center of the star $\delta\varphi$ behaves as
\begin{eqnarray}
\delta\varphi\approx c_1 r^{l+1}+c_2  r^{-l},
\end{eqnarray}
where $c_1$ and $c_2$ are constant of integrations.The scalar field doesn't have a singularity when $l=0$, but it does when $l>0$. By imposing $c_2=0$, the analytic solution of the Eq.(\ref{equper}) is
 \begin{eqnarray}
 \delta\varphi =b\;(C\,\hat{r})^{l+1} \left(1- C \, \hat{r}^2\right)^{-\frac{I \hat{\omega} }{2 \sqrt{C}}}{}_2F_1\left(m_{-}-\frac{I \hat{\omega}  }{2\sqrt{C}},m_{+}-\frac{ I \hat{\omega} }{2\sqrt{C}},l+\frac{3}{2},C \, \hat{r}^2\right)
 \end{eqnarray}
 where $\hat{\omega} =a \omega$ and $b$ is a constant of integration, ${}_2F_1$ is the hypergeometric function and
 \begin{eqnarray}
 m_\pm= \frac{1}{4} \left(2 l+3\pm\sqrt{9-24 \beta \kappa }\right).
 \end{eqnarray}
Outside the star, the solution of Eq.(\ref{equper}) is not evident and we cannot provide an analytic expression for the scalar field; nevertheless, an asymptotic behavior at infinity may be described as a linear combination of an outgoing and an ingoing wave 
\begin{eqnarray}
\delta\varphi\approx a_{in} e^{I\hat{\omega} \, \hat{r}^{*}}+a_{out} e^{-I\hat{\omega} \, \hat{r}^{*}},
\end{eqnarray}
where
the rescaled tortoise coordinate $\hat{r}^{*}$ is expressed as
\begin{eqnarray}
\hat{r}^{*}=\hat{r}+C \log(\hat{r}-C).
\end{eqnarray}
 Given our focus on gravastar quasi-normal modes, $\delta\varphi $ must manifest as a purely outgoing wave at infinity, without any ingoing wave component. However, for stable modes $Im(\hat{\omega} )>0$, the numerical solution suffer from numerical precision at infinity and we can not be sure if we have totally eliminated the ingoing wave. To overcome this problem and avoid error coming from ingiong wave, we use continued-fraction method \cite{leins1993}.\\

\begin{figure}[htb]
\centering
\includegraphics[width=0.45\textwidth, height=7cm]{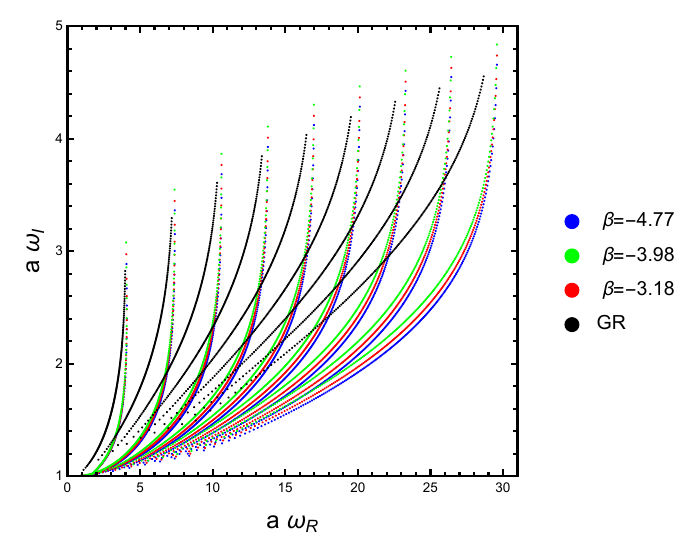} \hspace{1cm minus 0.25cm}
\includegraphics[width=0.45\textwidth, height=7cm]{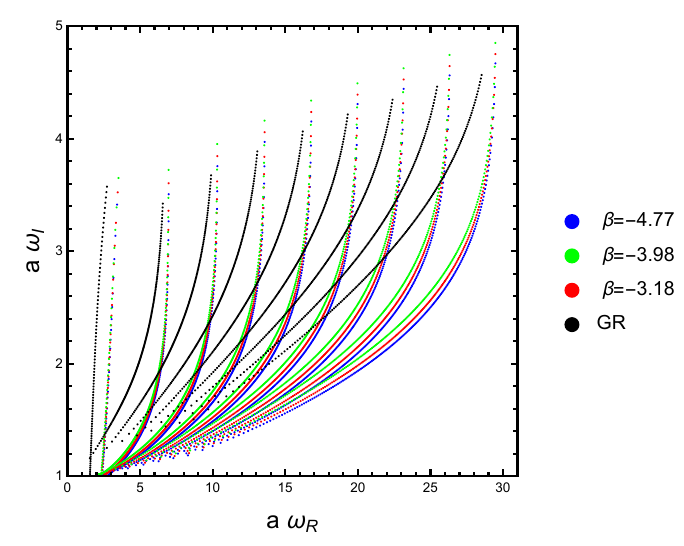}
\caption{\small Quasi normal modes of gravastars in schwarzschild black hole in scalar tensor theories and general relativity for the cases:  $l=0$ (left graph) and $l=2$ (right graph).}
\label{PGR}
\end{figure}

In a manner that is analogous to the situation of black holes, in which solutions are explored with an asymptotic behavior that simultaneously fulfills an outgoing wave at infinity and an ingoing wave at the horizon, we make use of an outgoing solution that contains a power series around the radius of the star
\begin{eqnarray}\label{expand}
\delta\varphi = e^{-I\hat{\omega} \, \hat{r}^{*}}\sum_{i=0}^{i=N}a_i \left(\frac{\hat{r}-1}{\hat{r}}\right)^{i},
\end{eqnarray}
where $a_i$ are complex constants and they are  related by the following  four-term recurrence relation
\begin{eqnarray}
a_{2}\alpha_1 + a_{1}\beta_{1} + a_{0}\gamma_{1}=0&\qquad n=1,\\
a_{n+1}\alpha_n + a_{n}\beta_{n} + a_{n-1}\gamma_{n}+ a_{n-2}\delta_{n}=0&\qquad n>1.
\end{eqnarray}
with
\begin{eqnarray}
\alpha_n = n (n+1)(1- C),\qquad \beta_n = n (n (3 C-2)-2 I \hat{\omega}  ),\nonumber\\
\gamma_n =n^2-n-l(l+1)- \left(3 n^2-3 n+1\right) C,\qquad \delta_n= (n-1)^2 C.
\end{eqnarray}
These expressions are true for any star with a schwarzschild exterior solution and they can written as a three-term recurrence relation \cite{leaver1990quasinormal}
\begin{eqnarray}\label{recurence}
a_{n+1}\hat{\alpha}_n + a_{n}\hat{\beta}_{n} + a_{n-1}\hat{\gamma}_{n}=0
\end{eqnarray}
where
\begin{eqnarray}
\hat{\alpha}_n = \alpha_n,\quad\hat{\beta}_n = \beta_n-\frac{\alpha_n \beta_{n-1}}{\hat{\gamma}_{n-1}},\quad\hat{\gamma}_n = \gamma_n-\frac{\delta_n \hat{\beta}_{n-1}}{\hat{\gamma}_{n-1}}.
\end{eqnarray}
By defining the recurrence relation
\begin{eqnarray}
a_{n+1}=R_n a_n,
\end{eqnarray}
then  substituting in (\ref{recurence}), we obtain
\begin{eqnarray}
R_n=-\frac{\hat{\gamma}_{n+1}}{\hat{\beta}_{n+1}+\hat{\alpha}_{n+1}R_{n+1}}.
\end{eqnarray}
 The quasi-normal frequencies $\hat{\omega} $ are those values for which the series in (\ref{expand}) exists and is finite, i.3. $a_n$ must be non-null and converge absolutely. For $n=0$, we have
 \begin{eqnarray}\label{a1a2}
 \frac{a_{1}}{a_{0}}=-\frac{\hat{\gamma}_{1}}{\hat{\beta}_{1}-\frac{\hat{\alpha}_{1}\hat{\gamma}_{2}}{\hat{\beta}_{2}-\frac{\hat{\alpha}_{2}\hat{\gamma}_{3}}{\hat{\beta}_{3}-\frac{\hat{\alpha}_{3}\hat{\gamma}_{4}}{\hat{\beta}_{4}-}}}}.
 \end{eqnarray}
where the left side of the above equation is determined from  the  continuity conditions at radius $[[\delta\varphi]]=0$ and $ [[\delta\varphi']]=0$,  as follow 
\begin{eqnarray}\label{ratio}
 \frac{a_{1}}{a_{0}}= 1+l+\frac{I \hat{\omega}  }{1-\sqrt{C}}+C(m_{+}-\frac{ I \hat{\omega} }{2\sqrt{C}})(m_{-}-\frac{ I \hat{\omega} }{2\sqrt{C}})\frac{{}_2F_1\left(1+m_{-}-\frac{ I \hat{\omega} }{2\sqrt{C}},1+m_{+}-\frac{ I \hat{\omega}  }{2\sqrt{C}},l+\frac{5}{2},C \right)}{(3 + 2 l){}_2F_1\left(m_{-}-\frac{ I \hat{\omega}  }{2\sqrt{C}},m_{+}-\frac{ I \hat{\omega}  }{2\sqrt{C}},l+\frac{3}{2},C\right)}.\nonumber\\
\end{eqnarray}
The nontrivial solutions of the equation (\ref{recurence}) exist only if Eq.(\ref{ratio}) is satisfied, where this is true for discrete value of $\hat{\omega} _n$. The identification of the QNMs, for a given values of the gravastars compactness $C$ and of the angular momentum integer $l$, has been reduced to a numerical issue, which is resolved using a root-finding algorithm in the complex plane. Since for any value chosen for $|R_N|<1$  will introduce a small error, which turns out to be negligible when $N$ is chosen to be large enough, we require the condition $R_N =0$ when $n$ is very large, in practice we choose $n=20$ \cite{Roussille:2023sdr}. Our results are showed in Figure.\ref{PGR}, where we compare the QNMs of our model with those in general relativity for different values of the compactness $C$ and the integer $l$. \\

In fact, we vary the compactness from $C=1/1000$ to $C=1$ for three values of the constant $\beta$, where $\beta$ is in the interval where spontaneous scalarizaition phenomena occurs. The values of the quasi normal modes collapse to the value $I$  as $C\rightarrow 1$, i.e. when the gravastar most closely approximates a BH for $l=0$ and $l=2$. In this limit, we see that the real part is very small compared to the imaginary part where the solution (\ref{ratio}) is not valid. Which means the QNMs of gravastars do not reduce to QNMs of black holes in both frameworks of general relativity and scalar tensor theories. The same conclusion is found in Ref.\cite{Pani:2009ss,Pani:2010em}. This can be seen by studying the interior solution in limit $C\rightarrow 1$ and $r\rightarrow 1$, i.e.
\begin{eqnarray}
\delta\varphi \approx \frac{e^{I\hat{\omega}  r^{*}}\Gamma\left(I \hat{\omega} \right)}{\Gamma\left(m_+ +I\, \hat{\omega} /2\right)\Gamma\left(m_{-} +I \,\hat{\omega} /2\right)}+\frac{e^{-I\hat{\omega}  r^{*}}\Gamma\left(-I \hat{\omega} \right)}{\Gamma\left(m_- -I \, \hat{\omega} /2\right)\Gamma\left(m_{+} -I \, \hat{\omega}/2 \right)}.
\end{eqnarray}
We see that the scalar field has ingoing and outgoing waves which are presented by the first and the second terms, respectively. If we wish to have only outgoing term at the exterior of the stars the first must vanish which happens at the poles of the gamma function in the denominator. Hence, the QNMs in the high compactness limit, for all values of $l$, are
\begin{eqnarray}
\hat{\omega} _{n'} = I2\,(\, n'+m_{+}),\quad n' =\{0,1,2,3....\}
\end{eqnarray}
which are purely imaginary number. Because of the term  $\Gamma\left(I \omega\right)$ in numerator, these modes do not exists in when $\beta=0$, i.e. the general relativity limit. The same result is obtained in Ref.\cite{choudhury2004quasinormal}. We Remarque that these frequencies are similar to the spectrum of QNMs for Schwarzshild black hole in high-damping regime \cite{nollert1999quasinormal,hod1998bohr}, where in this later the real part is a nonzero constant. These type of quasinormale modes are interpreted in the literature as  frequencies of excited levels of a black holes behaving as a damped harmonic oscillator \cite{Motl:2003cd}. Like in the previous works \cite{hod1998bohr,maggiore1994black,Motl:2003cd,nollert1999quasinormal}, we observe an equally spaced spectrum which will allow us to examine the propriety of quantum  black holes in semiclassical frame work via Bhor's correspondence. 
A transition from an exited black to a black hole might be an absorption of an energy equivalent to $\Delta M= \hbar  (\omega_{n}^{(0)}-\omega _{n-1}^{(0)})=4\pi T_H $, where $T_H = \hbar/(8\pi M)$ is Hawking temperature \cite{maggiore2008physical} and $\omega_{n}^{(0)}=\sqrt{(Re[\omega_{n}^{(0)}])^2+(Im[\omega_{n}^{(0)}])^2}$\cite{corda2012effective,corda2013effective}. If we follow the steps in these References, a gravastar in the limite of a black hole may also absorb energy as 
\begin{eqnarray}
\Delta M=8\pi T_H,
\end{eqnarray}
which differ  black hole by a factor of 2. And thus, according to second law of black hole thermodynamics the variation of the gravastar's era $\Delta A$ generated by $\Delta M$ is 
\begin{eqnarray}
\Delta A= -16\pi
\end{eqnarray}
once again the variation of the erea is different from those we get in a black hole by a factor of 2.  Note that this result is considerably different from Hod's conjuctor because we used Maggiore \cite{maggiore2008physical} rather than  Hod's one \cite{hod1998bohr}.

\section{Numerical analysis:}\label{sec4}

In the last section, we derived the solution  of a gravastar with a trivial scalar field where most of the solutions were analytic, e.i are reduced to general relativity. It is interesting to look into the physical characteristics of a non-vanishing scalar field at the background level and at the first-order perturbation. To do so, unlike in the case $\varphi=0$, we must give a form to the coupling function $A$ and we will consider the functions (\ref{models}) as examples. 
\subsection{Background solutions:}
 The presence of a  nontrivial scalar field will make the resolution of the differential equations (\ref{ef}), (\ref{eh}) and (\ref{ephi})  requires a numerical algorithm. The stability of the numerical solutions are ensured by  boundary conditions at the center of the star
 \begin{eqnarray}
 h(0)=1\quad \text{and}\quad \varphi'(0)=0,
 \end{eqnarray}
where the last condition will ensure the regularity of both scalar fields $\Phi$ and $\varphi$ at the center of the star. Additionally, it will guarantee the regularity of metric the Jordan frame at the center of the star, which is due to the fact that  the metric in Jordan frame is related to that in the Einstein frame by a conformal transformation defined in terms of  scalar field. However, these conditions are not sufficient to get asymptotically flatness at infinity which requires the boundary conditions
\begin{eqnarray}\label{infconditions}
f(\infty)=1\quad \text{and}\quad \varphi(\infty)=0,
\end{eqnarray}
and thus the metric in Einstein frame and Jordan frame become indistinguishable. In order to get the last condition we integrate the equations (\ref{ef}), (\ref{eh}) and (\ref{ephi}) using the specified starting condition
\begin{eqnarray}
\varphi(0)=\varphi_c,
\end{eqnarray}
where $\varphi_c$  is a real constant. In fact, using the transformation  (\ref{coordinatchangment}),  we integrate the background equation from $\hat{r}=0$ to $\hat{r}=1$ with $\rho=const$ and then from  $\hat{r}=1$ to $\hat{r}=\infty$ in the vaccum by taking into account the junction condition at $\hat{r}=1$ for the coupling functions (\ref{models}). During the integration, we need to give a numerical value to the energy density $\rho$ and the constant coupling $\beta$ wile $\varphi_c$ is chosen to  satisfy (\ref{infconditions}).

\begin{figure}[htb]
\centering
\includegraphics[width=0.4\textwidth, height=5cm]{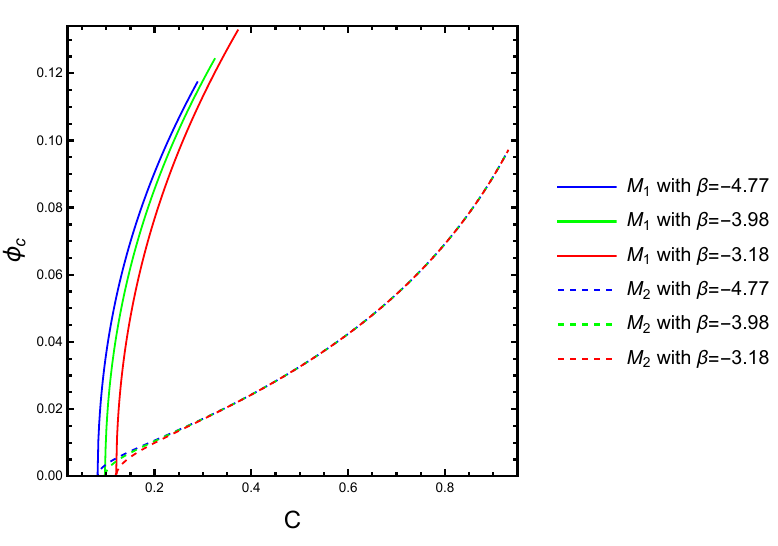} \hspace{1cm minus 0.25cm}
\includegraphics[width=0.4\textwidth, height=5cm]{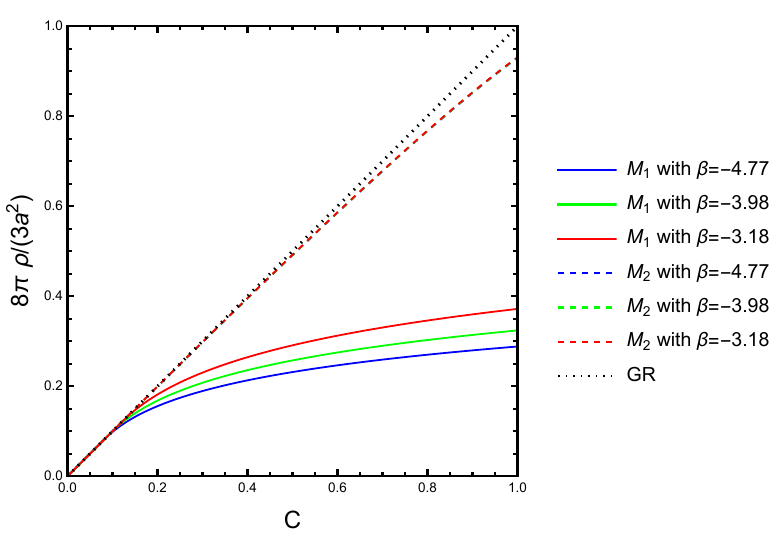} 
\caption{\small The variation of scalar field at the center of star (left graph) and the energy density  (right graph) as function as the compactness.}\label{CM}
\end{figure}

In the presence of the scalar field, the compactness $C$ can not obtained directly from the definition (\ref{grbacgroundre}), since we do not have the analytic expression of the metric. In this case, the behavior of the metric is 
\begin{eqnarray}
\lim_{\hat{r}\to\infty}  h(\hat{r})=1-\frac{C}{\hat{r}},
\end{eqnarray}
which equivalent to GR. Unlike in GR, the compactness of the star obtained from the above equation is different from the compactness derived from the energy density which is confirmed in the right of  Fig.\ref{CM} for both models $M_1$ and $M_2$. In Fig.\ref{CM}, we show the variation of the energy density of the star and central value of the scalar field w.r.t  compactness, where we can see that $\varphi_c$ is equal to zero in  interval $[0,C_c]$, where $C_c$ depends on the value of the constant $\beta$. Note the form of the coupling function do not contribute to the value of $C_c$ because the scalar field around this values is small and thus the both functions  are identical. Thus, in this interval the solutions are reduced to GR. By increasing the compactness, the central value of the scalar field is no longer zero and starts to increase at is illustrated in the left of  Fig.\ref{CM}. For larger gravarstar energy density in the presence of nontrivial scalar field leads to a decrease of the gravarstar compactness and  central value of the scalar field. \\

The variation of $\varphi_c$ and $\rho$ as function of $C$ depends on the value of $\beta$ for both models $M_1$ and $M_2$, but one can notice that the relations between these constants vary quantitatively. First, the relation $C-\rho$ in  model $M_2$ start to deviate from GR at high compactness while in model  $M_1$ is distinguishable from GR when $\varphi_c$ is not equal to zero. Second, the constant $\beta$ play an important role in the relations $C-\rho$ and $C-\varphi_c$ where by increasing $\beta$ the deviation from GR increase in the interval where $\varphi_c$ is not vanished. But, in the model $M_2$ we see that $\beta$ does not effect the  $\varphi_c-C$ only in small region close to $C_c$. We note that $C\approx\rho$  in $M_2$ at low compactness but significant deviation appears when $C\rightarrow 1$. Finally, it is expected that these differences will have impacts at the perturbed level which we will see in the next subsection. 

\begin{figure}[htb]
\centering
\includegraphics[width=0.4\textwidth, height=5cm]{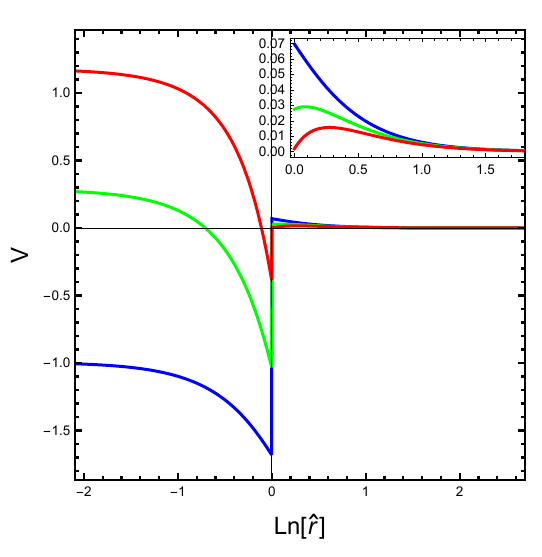} \hspace{1cm minus 0.25cm}
\includegraphics[width=0.4\textwidth, height=5cm]{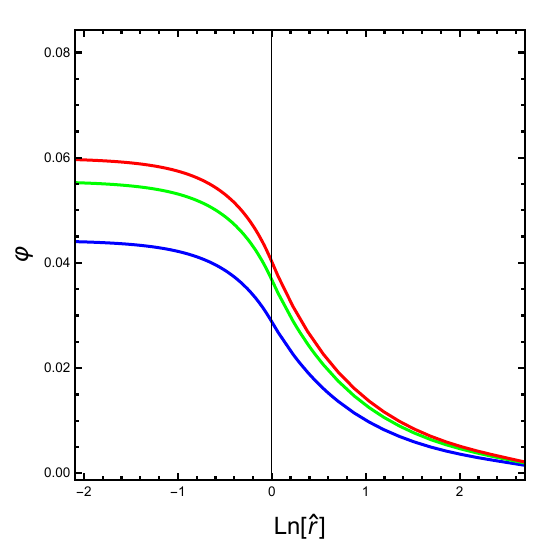} 
\caption{\small The The variation of the potential $V$ (in the right graph) and the scalar filed (in the left graph) $\varphi$ as function as $r$ for different values of  $\beta=\{-4.77\,(\text{Red\, line}),-3.98\,(\text{Green\, line}),-3.13\,(\text{Blue\, line})\}$.}
\label{V}
\end{figure}

\subsection{Scalar quasi normal modes:}
 
 In order to derive the quasi normal modes for non vanishing scalar field, we should integrate the equation (\ref{equper}) numerically  because  the scalar field and the metric do not have analytic expression in all spacetime. Here, we considered  the case $l=0$ where the perturbed metric and the perturbed scalar field are decoupled. In the case $\varphi\neq 0$, the Schrödinger-like differential equation has the same form as   (\ref{equper}) with the following potential
 \begin{eqnarray}
V(\hat{r})=\left\{
        \begin{array}{ll}
            f \left(\rho  \left(\frac{A^2-4 A \hat{r} A_{\varphi } \varphi '}{\kappa }-\frac{2 A^2 \hat{r}^2 \varphi '^2}{\kappa ^2}-A_{\varphi }^2-A A_{\varphi \varphi }\right)+\frac{1-h}{h\,\hat{r}^2}+\frac{2 \varphi '^2}{\kappa }\right),&\qquad \hat{r}<1,\\
            &\\
            f \left(\frac{1-h}{h\,\hat{r}^2}+\frac{2 \varphi '^2}{\kappa }\right),&\qquad \hat{r}>1.
        \end{array}
    \right.
\end{eqnarray}
 As an example, we plot $V(\hat{r})$ in Fig.\ref{V} for different values of $\beta$ using the model $M_1$. The method used in the last section to derive QNMs can not be applied in this case due to the numerical nature of the solutions for the background metric and the scalar field. Instead, the WKB approximation approach, as elucidated in Ref.\cite{kokkotas1992w}, where the author identified novel families of QNMs for relativistic neutron stars, can be employed to resolve our equation. Outside the star, the WKB approximation for the  equation (\ref{equper}) has the standard form
 \begin{eqnarray}
 \delta\varphi(\hat{\omega},\hat{r})= \frac{1}{Q(r)^{1/2}}\left(a(\omega)e^{-I\int^{\hat{r}^{*}}Q(r)dr}+ b(\omega)e^{I\int^{\hat{r}^{*}}Q(r)dr}\right),
 \end{eqnarray}
where $Q(r)=\sqrt{\hat{\omega}^2-V(r)}$. And  the interior solution $\delta\varphi_{int}$ is found from integrating inside the star and imposing the regularity condition $\delta\varphi\approx \hat{r}$. The junction conditions impose the continuity of the scalar field and it derivative trough the surface with the interior of the star. By doing so, we get the ratio
\begin{eqnarray}
\frac{b(\hat{\omega})}{a(\hat{\omega})}= e^{-2I\int^{1}Q(r)dr} \frac{Q(1)^2+\left(\frac{\delta\varphi_{int}'(1)}{\delta\varphi_{int}(1)}+\frac{Q'(1)}{2Q(1)}\right)^2}{\left(Q(1)+I\left(\frac{\delta\varphi_{int}'(1)}{\delta\varphi_{int}(1)}+\frac{Q'(1)}{2Q(1)}\right)\right)^2}.
\end{eqnarray}
in order to produce a pure outgoing wave function, this ratio must equal zero which happens  only for certain discrete values of $\hat{\omega}$. Nevertheless, this approach breaks down when $M \omega$ is very large and suffers from numerical errors when $M \omega\approx 1$ \cite{kokkotas1992w}. In order to overcome these issues, we employ the mixed numerical/WKB approach, where a numerical integration from infinity to the star's radius is performed as well as  the approximate relationship between the outgoing $\delta\varphi_{-}$ and ingoing $\delta\varphi_{+}$ waves functions,
\begin{eqnarray}
\frac{\delta\varphi_{+}'}{\delta\varphi_{+}}+\frac{\delta\varphi_{-}'}{\delta\varphi_{-}}=-\frac{Q_+'}{Q_+},
\end{eqnarray}
is used. From the continuity conditions, it follows that
\begin{eqnarray}\label{eqcount}
\frac{\delta\varphi_{+}'}{\delta\varphi_{+}}+\frac{\delta\varphi_{int}'}{\delta\varphi_{int}}=-\frac{Q_+'}{Q_+}.
\end{eqnarray}
The function $\delta\varphi_{+}$  is obtained by integrating inward from a large value of $\hat{r}$   to the stellar radius, using the ingoing wave function as the initial condition. This choice ensures that represents a purely ingoing wave. In contrast, obtaining $\delta\varphi_{-}$ from an outgoing wave function specified at infinity as the initial condition is more challenging, as ensuring its purity is not straightforward.
\begin{figure}[htb]
\centering
\includegraphics[width=0.4\textwidth, height=5cm]{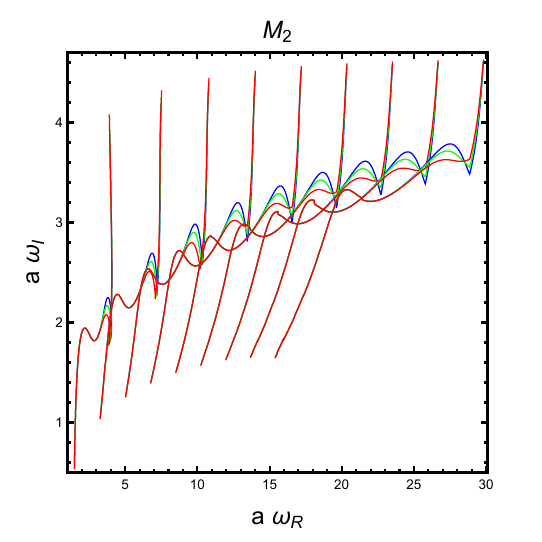} \hspace{1cm minus 0.25cm}
\includegraphics[width=0.4\textwidth, height=5cm]{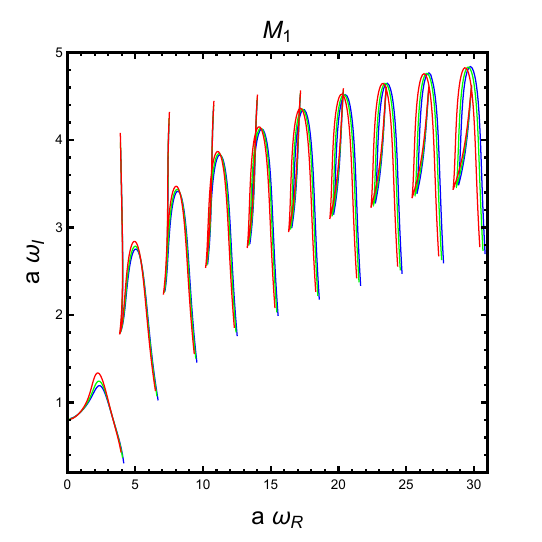} 
\caption{\small The first few quasinormale modes of scalar field perturbation equation for $\varphi\neq 0$ and for $l=0$. . In the left panel we QNMs of model $M_2$ is plotted for $\beta=\{-4.77(blue),-3.93(green),-3.18(red)\}$.  In the right panel we do the same for model $M_1$.}
\label{pc}
\end{figure}

To find the QNM frequencies, we fix the compactness $C$, the form of the coupling function  and  the constant $\beta$, and then we search for to  frequencies $\omega$ that verify the equation (\ref{eqcount}). Doing so, for each compactness, we get a complex frequency where its real  and imaginary components are illustrated in Fig.\ref{pc}. Starting from $C=0.001$ to $C=0.999$, we  produce the $\omega_R$-$\omega_I$ relation in the figure \ref{pc} for both models $M_1$ and $M_2$.  In the both models, we observe two part in the plots for each frequency, where the first part  is identical to GR and the second is not. This is due to manifestation of the scalar field which is the main reason for this deviation. Starting from $C=0.001$ to $C=0.999$, we calculate the $\omega_R$-$\omega_I$ relation shown in Figure \ref{pc}  for both models $M_1$  and $M_2$. In both models,  we observe that the plots have two distinct regions for each frequency. The first region is indistinguishable with the predictions of GR, while the second deviates considerably, which is caused principally by the presence the scalar field.

\section{Conclusion:}

In this present work, we have investigated the dynamics of gravitational  and scalar field in gravastars,  where  the conformal transformation has been utilized to couple the matter field with the metric. A numerical and analytic analysis has been performed to  describe the background and the scalar field. We have analyzed the scalar field equation of perturbation for the trivial  and nontrivial solutions. Using the countinued fraction method we compute QNMs for $\varphi=0$ and using WKB approximation for we compute QNMs for $\varphi\neq 0$.  We have provided key insights into the behavior of quasinormal modes and how the internal characteristics of gravastars affects them. \\

Our results have showed us that the scalar field and the metric  are significantly influenced by the choice of the form of the coupling function $A$ with $\beta$ considered as free parameter that control the interaction between different fields. This has lead to distinct observable signatures, highlighting the important role of the coupling function in shaping the physical feature of gravastars. We found that the gravastars is indistinguishable from GR case at small compactness and it starts to deviate from GR at high compactness. We found that the central value of the scalar field (the scalar field is not equal to zero in all the spacetime) is present when the compactness is bigger than a particular compactness. This behavior is  found in neutron stars and black holes \cite{doneva2020stability}. \\

 Through quasinormal modes, we expanded our understanding of the exotic compact object in alternative theories of gravity by studying the scalar field. The quasinormale frequencies provided us new behaviors by varying the compactness for both cases trivial and non trivial solutions. For the trivial solution, the form of the function $A$ does not impact the equations of perturbation  and thus the QNMs. However, when the scalar field is present we have observed  new feature of  QNMs where they have qualitative differences which depends on $A$. In fact, the model $M_1$ deviates considerably from GR and we found a mode which is absent in GR and in model $M_2$. \\
 
 In summary, our work contributes to the understanding of the characteristic of gravastars in frame of scalar tensor-theories, providing important insights into the configuration of these astrophysical objects. Extending our study to axial and polar perturbations will  gives us more information about the interior of gravastars. Finally, this study provides a path toward modeling and understanding gravastars in scalar-tensor theories, motivating us for further theoretical and observational exploration.
\appendix

\section{Coefficients:}\label{app.A}
The coefficients that appears in $\delta^2 S$ are given by:
\begin{eqnarray}
&&a_1= \frac{ \sqrt{f} r^2 \varphi'}{\sqrt{h}},\quad a_2= \frac{\kappa  }{2} r \sqrt{f},\quad a_3 =- \frac{ \kappa   }{2 \sqrt{h}}\sqrt{f}   r,\quad a_4=\text{C}_{\varphi }\text{C} \sqrt{f} r^2 \sqrt{h} \rho ,\quad a_5= \kappa\sqrt{f},\nonumber\\
&& a_6= -\frac{1}{4} \sqrt{f} \sqrt{h} \kappa  L -\frac{1}{2} \sqrt{f} \left( \sqrt{h} \kappa -h r^2 \rho \right) ,\quad a_7=\frac{\kappa}{4} \sqrt{f} \sqrt{h},\quad a_8=-2r^2 \varphi ',\nonumber\\
&& a_9=\kappa r,\quad a_{10}=-\frac{\kappa}{2}\sqrt{h}r,\quad  a_{11}=\frac{1}{4} \sqrt{f} \sqrt{h} \left( \kappa-\sqrt{h} \rho r^2\right),\quad a_{12}=\frac{\sqrt{f} r^2 \varphi ' }{ \sqrt{h}},\nonumber\\
&& a_{13}=-\text{C}\text{C}_{\varphi } \sqrt{f} \sqrt{h} r^2 \rho,\quad a_{14} =-\frac{\kappa   \left(r f'+2 f\right)}{4 \sqrt{f}} ,\quad a_{15}=\frac{\kappa }{2} \sqrt{f} \sqrt{h},\quad a_{16}=\frac{\sqrt{h} \kappa   r^2}{4 \sqrt{f}},\nonumber\\
&&  a_{17}=2\sqrt{f}  r \varphi ',\quad e_2= -e_3=- \frac{ \sqrt{f} r^2}{\sqrt{h}},\quad e_1 =-\sqrt{f} \sqrt{h}\left( L+\rho  r^2 \left(\text{C} \text{C}_{\varphi\varphi} +\text{C}_{\varphi}^2\right)\right).
\end{eqnarray}


\newpage

\bibliographystyle{ieeetr}
\bibliography{bibliography}

\end{document}